\documentclass[11pt,twoside,twocolumn]{article}
\usepackage{layout}
\usepackage{pgf}
\usepackage{amsmath,amssymb}
\usepackage{pgfpages}
\usepackage{verbatim}
\usepackage{subfigure}
\usepackage{abstract}
\usepackage{xspace}
\usepackage[labelfont=bf]{caption}
\usepackage[backend=biber,bibencoding=utf8,style=numeric-comp,doi=false,sorting=none,maxcitenames=1,maxbibnames=1]{biblatex} 
\renewbibmacro{in:}{%
  \ifentrytype{article}{}{%
  \printtext{\bibstring{in}\intitlepunct}}}

\bibliography{vlqbib} 
\usepackage{latexsym}
\usepackage{feynmf}
\usepackage{mycommands,mystyle}
\usepackage{lineno}
\usepackage[small,compact]{titlesec}
\usepackage{multicol}
\usepackage[
pdftitle={Proceedings LP2025 -- Searches for VLQs and LQs from the ATLAS Experiment -- Elin Bergeaas Kuutmann},
pdfauthor={Elin Bergeaas Kuutmann},
pdfsubject={Proceedings LP2025},
pdfkeywords={Proceedings LP2025, Searches for VLQs and LQs from the ATLAS Experiment, Elin Bergeaas Kuutmann},
colorlinks=true,
pdfstartview=FitV,
linkcolor=blue,
citecolor=blue,
urlcolor=blue,
bookmarks=true,
bookmarksnumbered=true,
bookmarksopen=true,
bookmarksopenlevel=4
]{hyperref}
\usepackage{fancyhdr}

\fancypagestyle{plain}{%
  \fancyhead{} %
  
}
\newcommand{\kommentar}[1]{}
 % ...
\kommentar{%
   the text in the brackets is interpreted as comment...
  }%

\title{Searches for VLQs and LQs from the ATLAS Experiment}
\author{Elin Bergeaas Kuutmann (Uppsala University), on behalf of the ATLAS Collaboration}

\makeatletter
\def\@maketitle{%
  \newpage
  \null
  \vskip 2em%
  \begin{center}%
  \let \footnote \thanks
    {\LARGE \@title \par}%
    \vskip 1em%
    {\large Presented at the 32nd International Symposium on Lepton Photon Interactions at High Energies, Madison, Wisconsin, USA, August 25-29, 2025\par}%
    \vskip 1.5em%
    {\large
      \lineskip .5em%
      \begin{tabular}[t]{c}%
        \@author
      \end{tabular}\par}%
    \vskip 1em%
    {\large \@date}%
  \end{center}%
  \par
  \vskip 1.5em}
\makeatother

\date{\today}

\makeindex
\begin{document}

\setlength{\parindent}{0pt}  
\setlength{\parskip}{0.8ex}

\hyphenation{ATLAS Upp-sala an-aly-ses}

\pagenumbering{arabic}

\setcounter{page}{1}

\thispagestyle{empty}

  \twocolumn[
\maketitle 
\begin{onecolabstract}
  The Standard Model of particle physics explains many natural phenomena yet remains incomplete. Vectorlike quarks and leptoquarks lie at the heart of many extensions to the Standard Model seeking to address the hierarchy problem,  or the flavour sector anomalies. These proceedings present the new results from searches with the ATLAS detector at the LHC\footnotemark.
\end{onecolabstract}
  ]

  \footnotetext{\copyright Copyright 2025 CERN for the benefit of the ATLAS Collaboration. CC-BY-4.0 license.}

\section{Physics beyond the Standard Model}

Could more generations of matter exist beyond the three known ones from the Standard Model (SM)?
The answer is no, at least if we require light neutrinos (shown by the LEP experiments~\cite{ALEPH:2005ab}).
A fourth generation of fermions that couple to the Higgs boson has also been excluded by electroweak (EW) precision measurements~\cite{Eberhardt:2012gv}.
Hence, if new matter-like particles exist, they must be different from the ones in the SM. Vector-like quarks (VLQs), vector-like leptons (VLLs) and leptoquarks (LQs) are possible, not yet excluded, extensions to the SM particle content.

In these proceedings, searches for VLQs, VLLs and LQs in the ATLAS experiment~\cite{atlas_detector_paper} are discussed. 
All results are based on the full Run 2 dataset at $\sqrt{s}=13$~\TeV, unless otherwise stated. All limits are given at the 95\% confidence level. 

\section{Vector-like quarks}
\label{sec:VLQ}

VLQs are hypothetical particles beyond the SM 
which, like quarks, carry color charge and have spin $\frac{1}{2}$.
Unlike quarks, their right and left components have the same quantum numbers, i.e.~they are ``vector-like'', not chiral.
A consequence of this is that they don't have to acquire mass via the Higgs mechanism. 
VLQs arise in for example Composite Higgs models and little Higgs models and can offer an explanation to the Higgs fine-tuning problem~\cite{Perelstein:2003wd,Matsedonskyi:2012ym}. 

\begin{figure}
  \subfigure[$Y$ or $T \rightarrow Wb$]{
\label{F:VLQsingleprodYTtoWb}
\raisebox{1.2ex}{\includegraphics[width=0.45\linewidth]{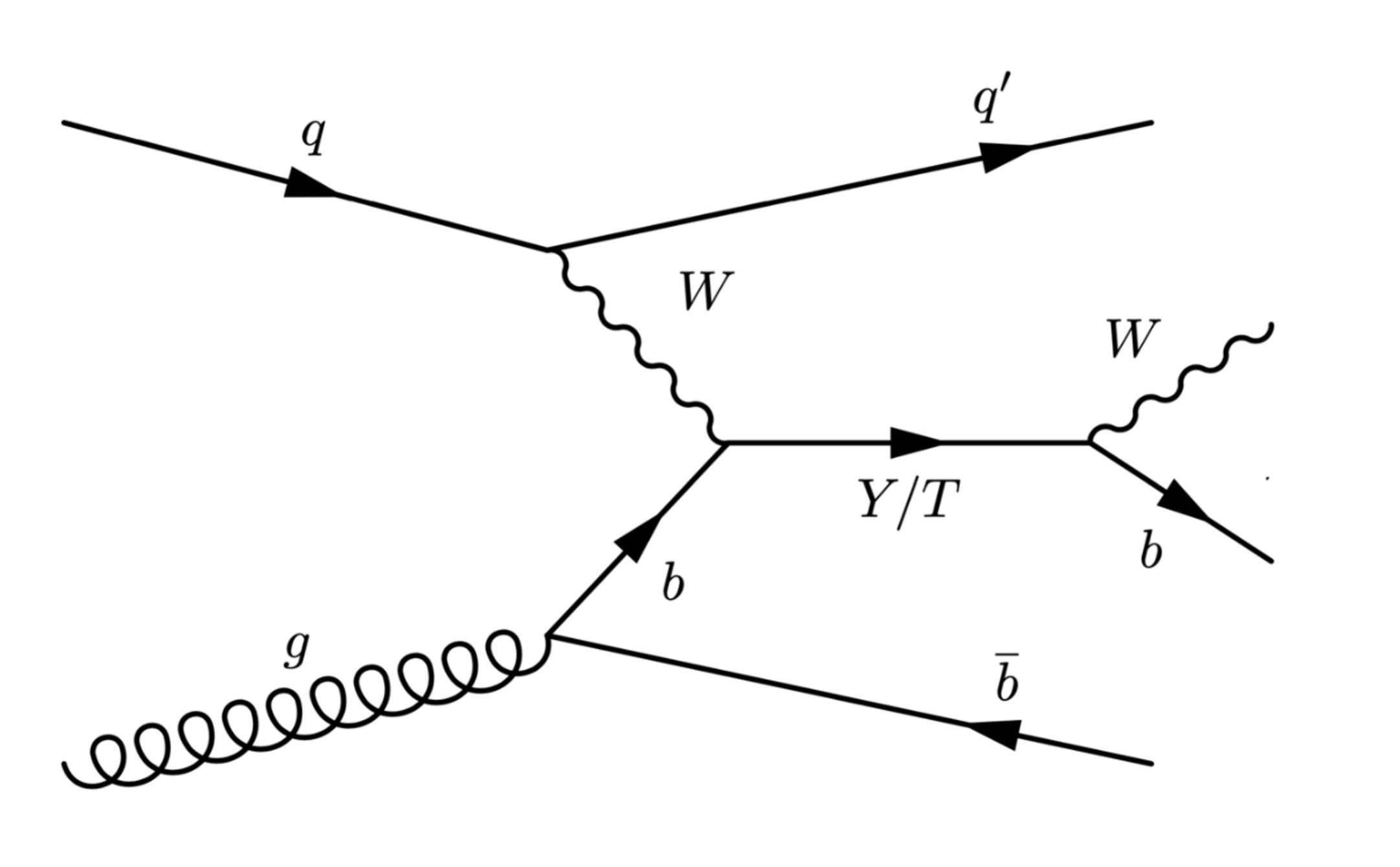} }}
\subfigure[$T \rightarrow H$ or $Z$ and $t$]{
\label{F:VLQsingleprodTtoHZt}
\includegraphics[width=0.45\linewidth]{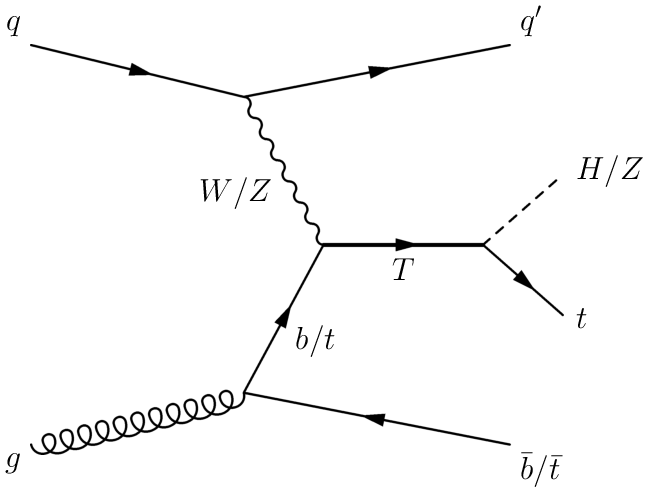} }
\caption{Examples of single production and decay of VLQs.
  \label{fig:VLQprod}}
\end{figure}

VLQs can be produced in pairs (through QCD processes, not discussed here) or singly through EW processes, which makes the production cross section and branching ratios model-dependent. Two examples of single VLQ production are shown in 
Figure~\ref{fig:VLQprod}.

By convention~\cite{Aguilar-Saavedra:2013qpa}, the VLQ $T$ has electric charge $+\frac{2}{3}$, $B$ has charge $-\frac{1}{3}$, and $Y$ has charge $-\frac{4}{3}$. Depending on the model under consideration, there could also be VLQs with other charges such as $X_{5/3}$, $Y_{8/3}$, etc.
VLQs can be singlet, doublet or triplet under SU(2), which has consequences for e.g. their couplings and branching ratios. 
The coupling constant $\kappa$ characterizes the VLQ coupling, 
e.g.~$YWb$ of Figure~\ref{F:VLQsingleprodYTtoWb}. 
$\kappa$ is proportional to the decay width $\Gamma$ and dependent on the underlying model.

All analyses considered here assume that the VLQ decays into SM bosons and a third-generation quark.
A summary of the current best exclusion limits from all VLQ searches in ATLAS can be found in the VLQ summary plot~\cite{ATL-PHYS-PUB-2025-030}.

\subsection{Search for singly produced $T/Y \rightarrow Wb$ in the one lepton channel} 

\begin{figure}[h!]
\begin{center}
\includegraphics[width=0.95\linewidth]{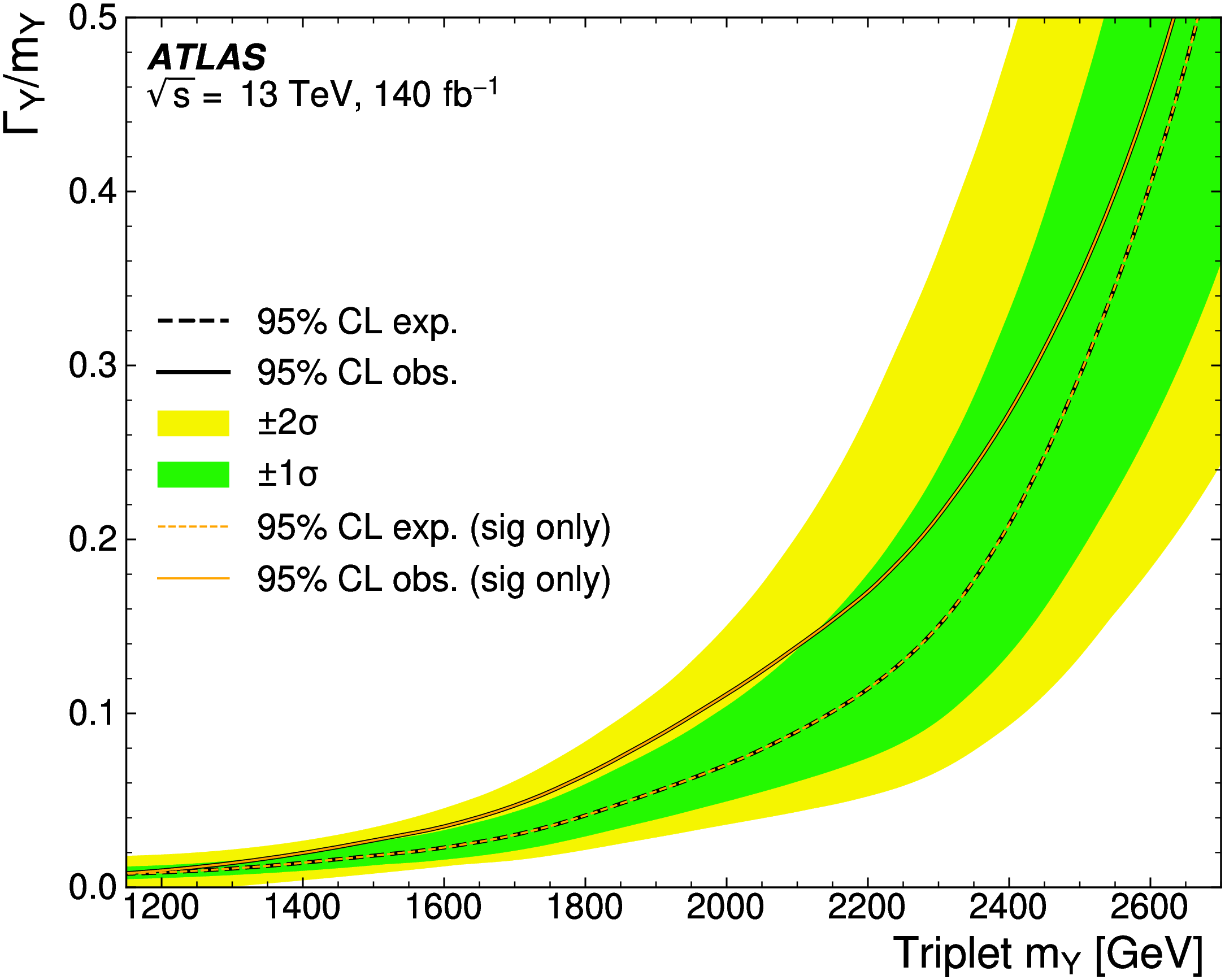} 
\caption{Exclusion limits of $Y \rightarrow Wb (1\ell)$ in the width -- mass plane. From~\cite{EXOT-2018-60}. \label{fig:YtoWbOneLLimit}}
\end{center}
\end{figure}

A recent ATLAS paper describes the search for an SU(2) singlet $T$ or a $Y$ from the SU(2) $(T, B, Y)$ triplet in the final state of $W+b \rightarrow e/\mu +\nu + b$~\cite{EXOT-2018-60}.
The process is shown in Figure~\ref{F:VLQsingleprodYTtoWb}. 
In this analysis, single lepton triggers are used to look for a signature of a lepton, $\ell$ (an electron or a muon).
In addition, candidate events are required to contain missing transverse energy (\met), one hard $b$-jet and a forward jet from the light quark $q'$ in Figure~\ref{F:VLQsingleprodYTtoWb}.
The reconstructed mass of the VLQ, $m_{\mathrm{VLQ}}$, comes from the lepton, \met and the $b$-jet. Limits are set in terms of the coupling constant $\kappa$: 
For a singlet $T$, the $\kappa$ limit ranges between 0.22 and 0.52 for $m_T$ between 1150 and 2300 \GeV. For a $Y$ from the triplet, the limit is found between 0.14 and 0.46 for $m_Y$ from 1150 to 2600 \GeV. 
The limits are also given in terms of decay width and VLQ mass, as shown in Figure~\ref{fig:YtoWbOneLLimit}, for the $Y$ case. 

The production of $Y$, as shown in Fig.~\ref{F:VLQsingleprodYTtoWb}, can interfere with the SM background. This effect was explicitly investigated and found to be negligible, since the interference-included (black) and signal-only (orange) exclusion curves in Figure~\ref{fig:YtoWbOneLLimit} are nearly indistinguishable. 
In the $T$ case, there is no interference with the SM.

The limits reinforce and extend the mass coverage of the singly produced VLQ upper cross section limits from ATLAS.

\subsection{Search for singly produced $T/Y \rightarrow Wb$ in the all-hadronic channel}

Complementary to the previous analysis, also a search for the $T/Y \rightarrow Wb$ process in the all-hadronic final state has been performed~\cite{EXOT-2022-43}, where limits are set on an SU(2) singlet $T$, and a $Y$ from the $(B, Y)$ doublet. 
A large-$R$ jet trigger is used to select candidate events, with a signature of a $W$-tagged large-$R$ jet and an additional small-$R$ $b$-jet. The candidate VLQ mass is reconstructed as the invariant mass of the $W$-tagged and the $b$-tagged jets.

\begin{figure}[htb]
\begin{center}
\includegraphics[width=0.95\linewidth]{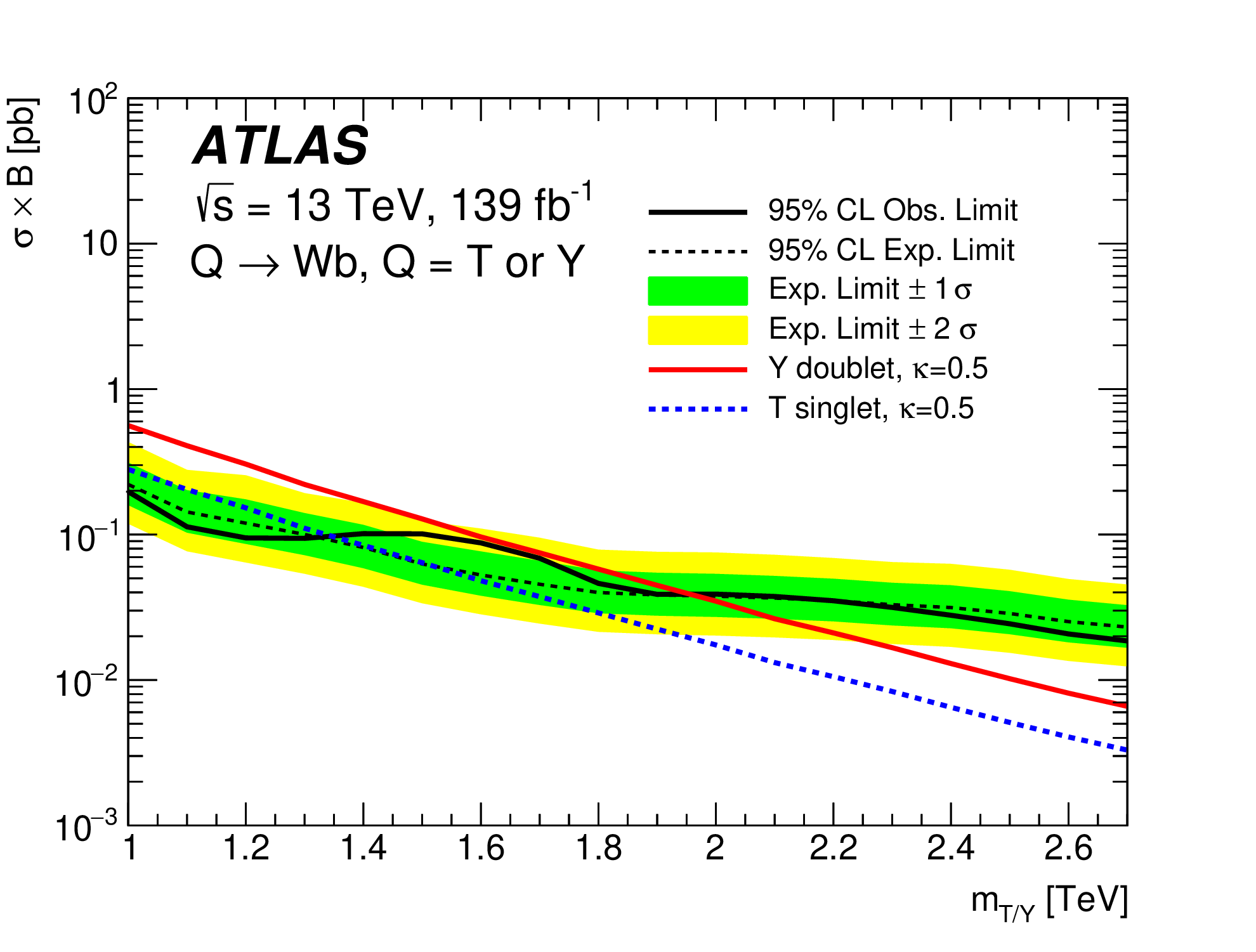} 
\caption{Upper cross section limits as a function of the VLQ mass for the $T/Y \rightarrow Wb$ search in the all-hadronic channel. From~\cite{EXOT-2022-43}. \label{fig:YtoWbAllHadLLimit}}
\end{center}
\end{figure}

Limits are set depending on the assumed $\kappa$ coupling value, where the observed lower mass limits for a $Y$ quark were $2.0$~\TeV\ for $\kappa = 0.5$ and $2.4$~\TeV\ for $\kappa = 0.7$. For the $T$ quark the limits are $1.4$~\TeV\ for $\kappa = 0.5$ and 1.9 \TeV\ for $\kappa$ = 0.7.
Figure~\ref{fig:YtoWbAllHadLLimit} shows the upper cross section limit as a function of VLQ mass, with the predicted cross sections for the $Y$ doublet and the $T$ singlet, assuming $\kappa = 0.5$.

The analysis improves the mass limits on $Y$ in the $(B,Y)$ doublet w.r.t.~previous ATLAS results, and the search regions for $Y$ and the $T$ singlet have been extended.

\subsection{Search for $t$ in a mono-top and \MET final state}

A search for events with a single high-energy boosted top quark (a mono-top) in addition to large amounts of \met has been performed in the ATLAS experiment~\cite{EXOT-2022-40}.
This final state can be interpreted as candidates for singly produced $T$ decaying into $Zt$, where the $Z$ decays into two neutrinos, detected as \met. The process is shown in Figure~\ref{F:VLQsingleprodTtoHZt}. 
For this analysis, $\met$ triggers are used and the selected events must contain a top-tagged large-$R$ jet in addition to a large amount of \met. 
The XGBoost classifier~\cite{Chen_2016} is used to separate the background from signal. 

\begin{figure}[htb]
\begin{center}
\includegraphics[width=0.95\linewidth]{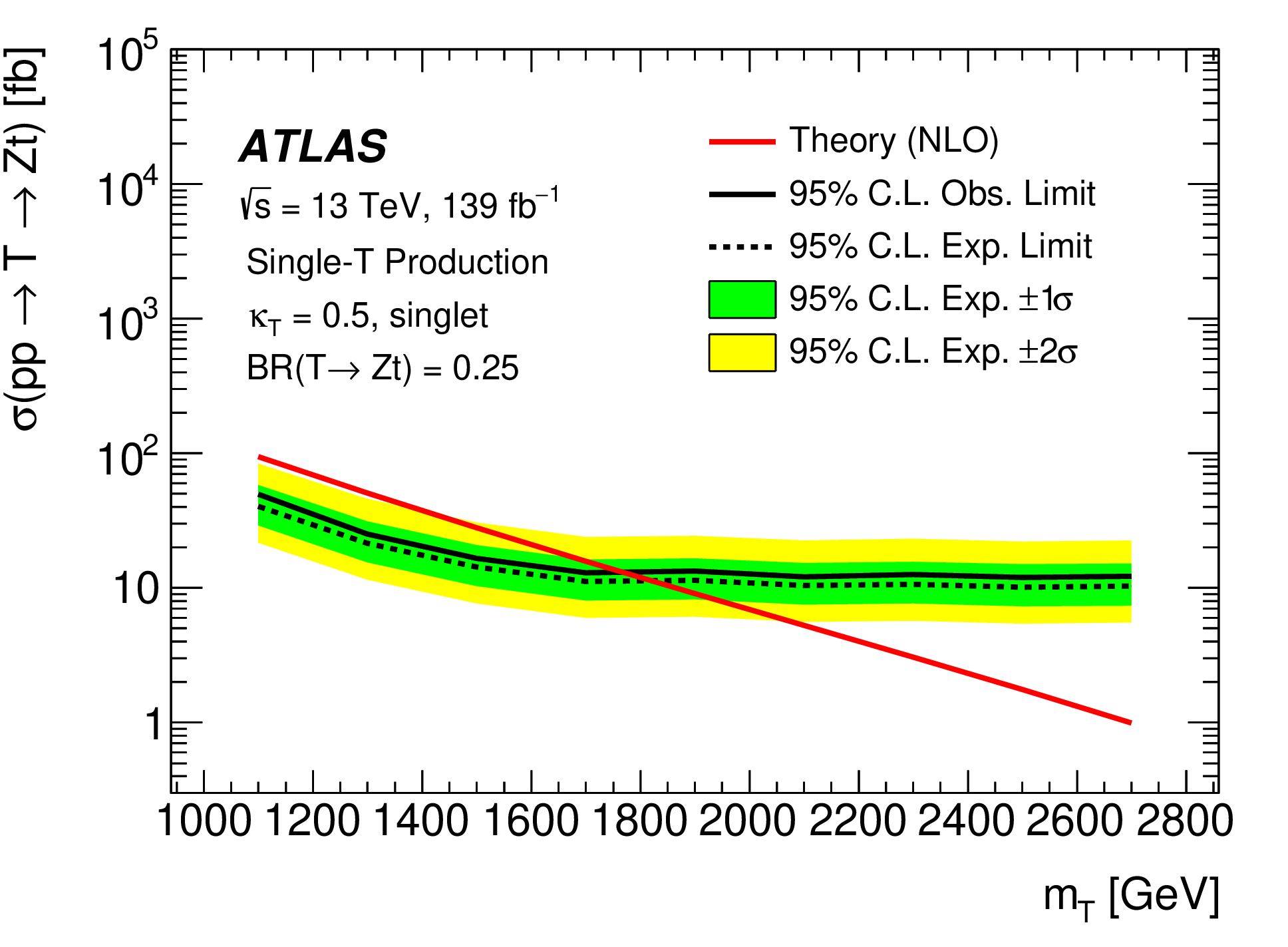} 
\caption{Upper cross section limits on $T \rightarrow Z(\nu\nu)t$ as a function of mass. From~\cite{EXOT-2022-40}. \label{fig:TtoZtLimits}}
\end{center}
\end{figure}

The lower mass limit on an SU(2) singlet $T$ mass was found to be $1.8$~\TeV\ for a coupling parameter value of $\kappa_T = 0.5$ and under the assumption that the branching ratio fulfills $\mathrm{BR}(T \rightarrow Zt) = 25\%$.
The upper cross section limit as a function of mass is shown in Figure~\ref{fig:TtoZtLimits}.

\subsection{Single $T$ combination}
Recent ATLAS searches for singly produced $T$ decaying into $t$ and $H$ or $Z$, including the mono-top search from the previous section, have been combined~\cite{EXOT-2021-02}. 

\begin{figure}
\begin{center}
\includegraphics[width=0.95\linewidth]{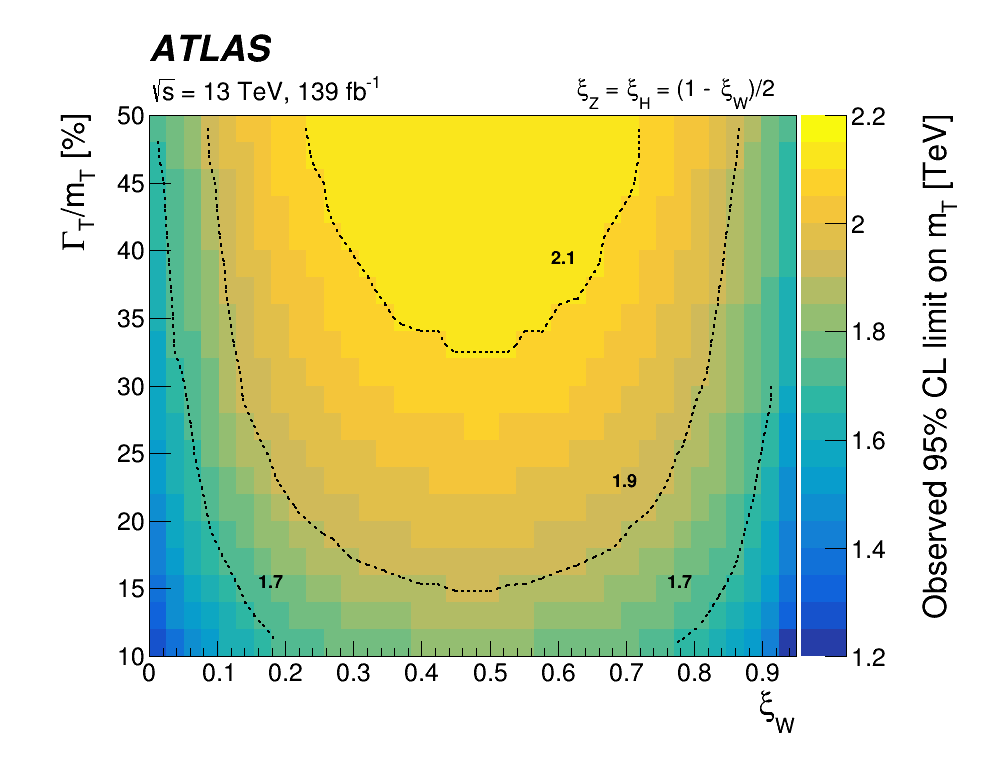} 
\caption{Lower mass limit on $T$ as a function of the relative coupling $\xi_W$ (which is $BR(T\rightarrow Wb)$ for large $m_T$). From~\cite{EXOT-2021-02}. \label{fig:SingleTcomb}}
\end{center}
\end{figure}

Results are given as a function of the decay width $\Gamma$ and the branching ratio into $W, H, Z$. In Figure~\ref{fig:SingleTcomb} the observed lower mass limit is parametrised as a function of the relative coupling $\xi_W$, 
which corresponds to the branching ratio into $Wb$ for large $T$ masses.
In this plot, $\xi_W=0.5$ corresponds to the SU(2) singlet $T$ and $\xi_W=0.0$ corresponds to the SU(2) doublet $T$. 
This holds under the assumption that the $T$ only decays into a $H$, $Z$ or $W$, and a third generation quark.

\section{Leptoquarks}
\label{sec:LQ}

\begin{figure}
\subfigure[Leading order resonant LQ production.]{
\label{F:LQprodThree}
\includegraphics[width=0.4\linewidth]{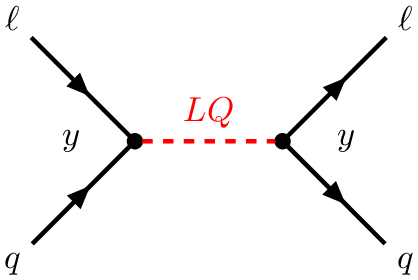} } 
\hspace{1em}
\subfigure[Next-to-leading order resonant LQ production.]{
\label{F:LQprodFour}
\includegraphics[width=0.4\linewidth]{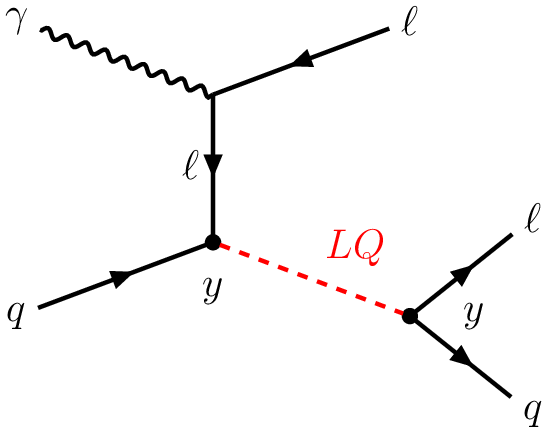} }
\caption{Examples of single resonant production of a LQ and its subsequent decay into a lepton and a quark. The parameter $y$ 
  is the LQ Yukawa coupling to leptons and quarks. 
  \label{fig:LQprod}}
\end{figure}

LQs can arise in GUT models with extended gauge groups. They are hypothetical particles which carry both lepton and baryon numbers, as well as electric charge and color. They are expected to decay into a lepton-quark pair.
The models usually predict them to be scalars (spin 0) or vectors (spin 1).
If they exist, they would lead to lepton flavor universality violations.
Like VLQs, they can be  produced in pair (through QCD processes) or singly, which is model-dependent. Examples of single production can be seen in Figure~\ref{fig:LQprod}. 
In these proceedings, one search for single resonant production of LQs is covered, which is complementary to previous ATLAS searches for pair-produced LQs. A recent search for vector-like leptons (VLLs) decaying into vector-like quarks is also discussed.
An overview of the current best exclusion limits of LQs can be found in the LQ summary plot~\cite{ATL-PHYS-PUB-2025-013}. 

\subsection{Search for a lepton-jet resonance}
The ATLAS collaboration has performed a search for a single resonant production of a scalar LQ, as shown in Figure~\ref{fig:LQprod}, which couples to charged leptons and down-type quarks~\cite{EXOT-2024-12}. The full Run~2 data set as well as 55~$\ifb$ of Run 3 data have been used, and this search is complementary to earlier pair-produced LQ searches. 

Lower limits on the mass are given depending on the coupling $y$. The lower mass limit is 3.4 \TeV\ for $y_{de}=1.0$ (assuming decay into an electron and a $d$ quark), 4.3 \TeV\ for $y_{s\mu} =3.5$, 3.1 \TeV\ for $y_{be}=3.5$ and 2.8 \TeV\ for $y_{b\mu}=3.5$. Limits for the latter case, stated as $y_{b\mu}$ versus the LQ mass are shown in Figure~\ref{fig:LeptonJetLimits}. 
\begin{figure}
\begin{center}
\includegraphics[width=0.95\linewidth]{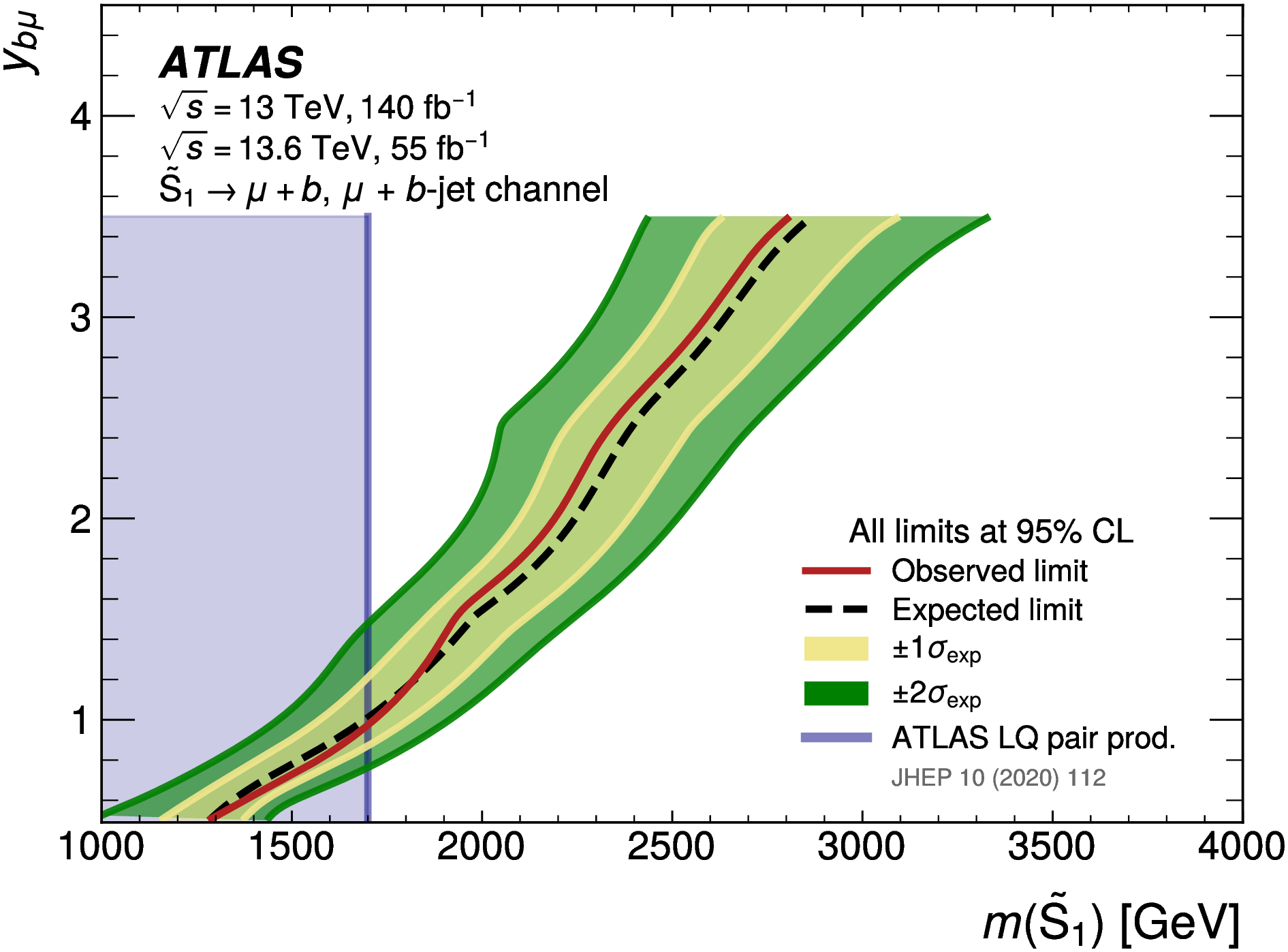} 
\caption{Lower mass limits for a scalar leptoquark decaying into a muon and a $b$-quark. From~\cite{EXOT-2024-12}. \label{fig:LeptonJetLimits}}
\end{center}
\end{figure}
These results extend previous mass limits for LQs.

\subsection{Vector-like leptons decaying to a LQ}

If you can have vector-like quarks, why not leptons? In a recent ATLAS analysis, a search for VLL is performed, under the hypothesis that they decay into vector LQs, which decay into a $\tau$ lepton or a neutrino and a $b$-quark or a top quark~\cite{EXOT-2022-27}.
The production mode and decay are shown in Figure~\ref{fig:VLLFeynman}.
Here $N$ is a VLL which is electrically neutral like the neutrino, and the VLL $E$ is charged like the electron. The vector LQ is called $U_1$ in this model, 
\begin{figure}
\begin{center}
\includegraphics[width=0.95\linewidth]{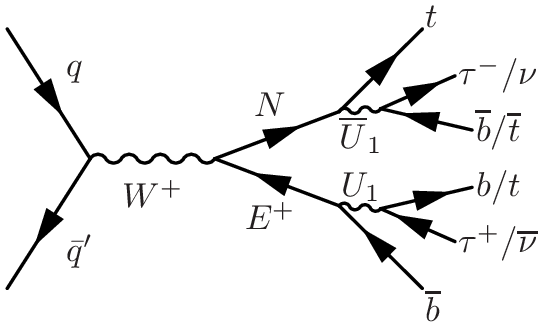} 
\caption{Pair-production of VLLs ($E$ and $N$ and their decay into vector LQs $U_1$ which decay into third-generation leptons and quarks. From~\cite{EXOT-2022-27}. \label{fig:VLLFeynman}}
\end{center}
\end{figure}
an ultraviolet-complete extension to the SM, called ``4321''~\cite{DiLuzio:2018zxy}.%

\begin{figure}
\begin{center}
\includegraphics[width=0.95\linewidth]{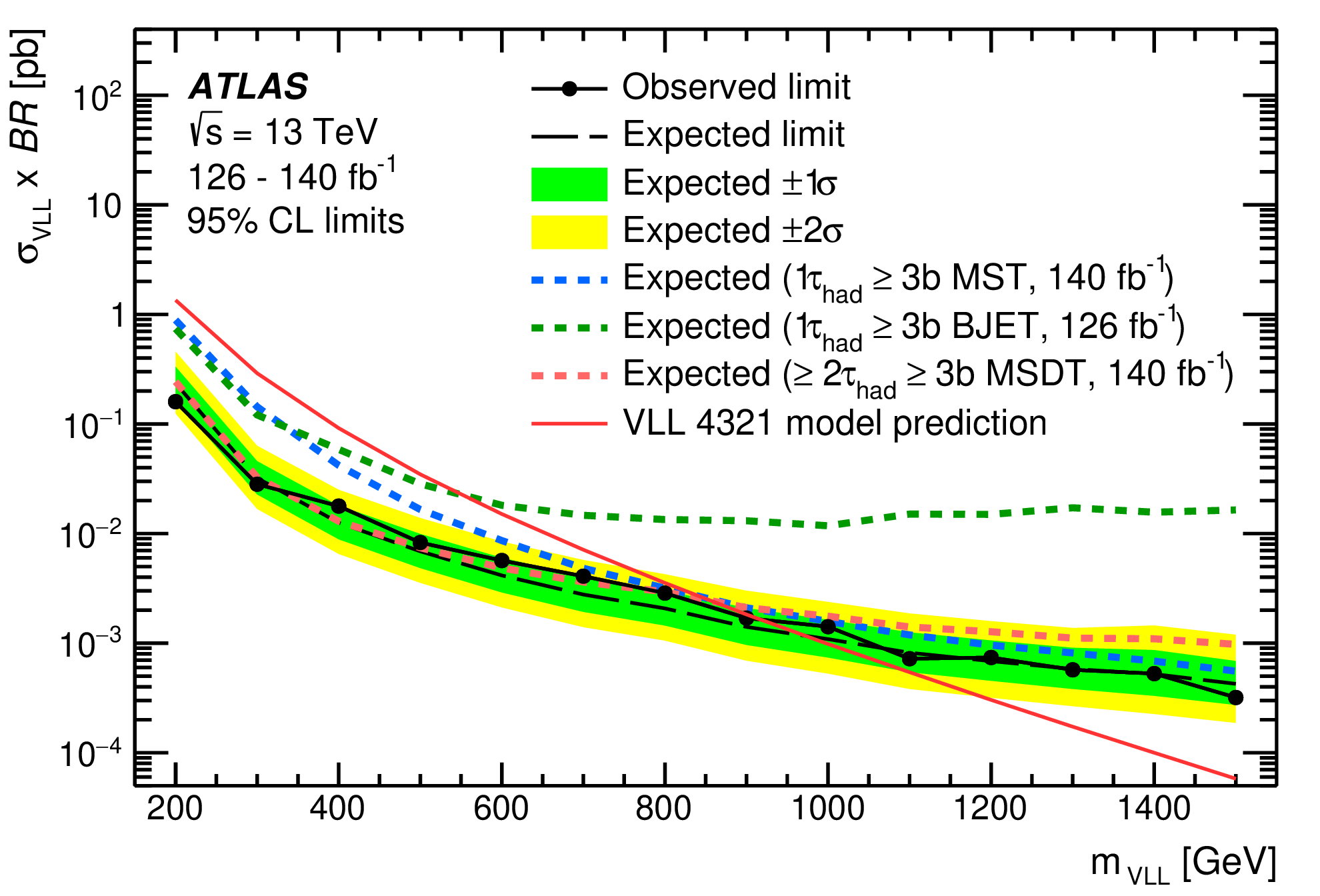} 
\caption{Upper cross section limit as a function of VLL mass in the ``4321'' model. From~\cite{EXOT-2022-27}. \label{fig:VLLLimits}}
\end{center}
\end{figure}

The analysis excludes masses of the VLL from 200 to 910 \GeV\ for the cross-section predicted by the ``4321'' model, as shown in Figure~\ref{fig:VLLLimits}.

\section{Summary and conclusions}
VLQs, VLLs and LQs are possible new particles beyond the Standard Model which have not yet been excluded by direct observations or precision tests. 
ATLAS has extensive programmes to search for these particles in a wide range of channels.
The recent results, mostly searches in single production channels, 
expand the exclusion limits in both mass and couplings 
and improve our knowledge of what properties heavy particles in scenarios beyond the Standard Model could have. 

\subsection*{Acknowledgments}

The author wishes to thank the Carl Trygger Foundation, Sweden, for generous travel funds under the grant CTS 22:2312.

  \printbibliography

\end{document}

\typeout{get arXiv to do 4 passes: Label(s) may have changed. Rerun}